\documentclass[conference,a4paper]{IEEEtran}
\IEEEoverridecommandlockouts
\usepackage{cite}
\usepackage{amsmath,amssymb,amsfonts}
\usepackage{graphicx}
\usepackage{textcomp}
\usepackage{xcolor}
\usepackage{booktabs}
\usepackage{caption}
\usepackage{url}
\usepackage[hidelinks]{hyperref}
\def\BibTeX{{\rm B\kern-.05em{\sc i\kern-.025em b}\kern-.08em
    T\kern-.1667em\lower.7ex\hbox{E}\kern-.125emX}}
\begin{document}

\title{Validating Threat Modeling Results with the Help of Vulnerable Test Applications}

\author{%
\IEEEauthorblockN{Oleksandr Adamov\IEEEauthorrefmark{1}, Davide Fucci\IEEEauthorrefmark{1}, Felix Viktor Jedrzejewski\IEEEauthorrefmark{1}, Ricardo Britto\IEEEauthorrefmark{2}, Nishrith Saini\IEEEauthorrefmark{3}}
\IEEEauthorblockA{\IEEEauthorrefmark{1}Blekinge Institute of Technology, Karlskrona, Sweden\\
\{oleksandr.adamov,davide.fucci,felix.jedrzejewski\}@bth.se}
\IEEEauthorblockA{\IEEEauthorrefmark{2}Ericsson, Kista, Sweden\\
ricardo.britto@ericsson.com}
\IEEEauthorblockA{\IEEEauthorrefmark{3}Ericsson, Karlskrona, Sweden\\
nishrith.saini@ericsson.com}
}

\maketitle

\begin{abstract}
Validating threat modeling results remains difficult because completeness is hard to judge without an external oracle. Existing studies often rely on expert-produced reference models and other human baselines, but these can contain omissions or disagreements. This paper evaluates a complementary, vulnerability-grounded validation approach. We apply threat modeling to intentionally vulnerable applications with a known vulnerability set to measure the number of related vulnerabilities that can be discovered. We compare ThreMoLIA, an LLM-assisted threat modeling solution developed by our team, with the Microsoft Threat Modeling Tool (MTMT) across two vulnerable applications: AzureGoat and the Vulnerable Bank Application (VulnBank). The inputs to both tools are limited to architecture, data flow diagrams, and their descriptions. The results show that ThreMoLIA achieved higher vulnerability coverage on both systems. We show that vulnerable test applications provide a practical benchmark for assessing threat coverage and complement expert-based validation.
\end{abstract}

\begin{IEEEkeywords}
Threat modeling, LLM, STRIDE, vulnerability, weakness, benchmark, ThreMoLIA, telecommunications.
\end{IEEEkeywords}

\section{Introduction}
Threat modeling is a well-established security-by-design practice for identifying potential threats early, before implementation defects become costly to fix \cite{shostack2014threat, xiong2019threat}. At the same time, the quality of a threat model is difficult to assess objectively. Systematic reviews report that threat modeling remains largely manual and that validation assurance remains limited \cite{xiong2019threat, yskout2020threat}. In practice, evaluations commonly compare the produced threat model against an expert's reference model or another human baseline. This is useful, but it also inherits the limitations of the reference model itself: experts may disagree, work at different abstraction levels, or simply miss relevant threats.

This paper studies a complementary validation approach based on vulnerable test applications, where a known set of vulnerabilities serves as an oracle to assess how well a threat model covers weaknesses and vulnerabilities that are actually present in the system.

We use this approach to compare ThreMoLIA and the Microsoft Threat Modeling Tool (MTMT). ThreMoLIA is an LLM-assisted approach for threat modeling of LLM-integrated applications that relies on project documentation, high-level architecture, and data flow diagrams. ThreMoLIA uses threat frameworks for traditional applications, such as STRIDE\footnote{STRIDE: \emph{Spoofing, Tampering, Repudiation, Information Disclosure, Denial of Service, and Elevation of Privilege}; https://learn.microsoft.com/en-us/azure/security/develop/threat-modeling-tool.}, 
LINDDUN\footnote{LINDDUN: \emph{Linkability, Identifiability, Non-repudiation, Detectability, Disclosure of information, Unawareness, and Non-compliance}, https://linddun.org/.}, 
OWASP Top 10\footnote{The OWASP Top 10 is a standard awareness document covering the most critical web application security risks, https://owasp.org/www-project-top-ten/.}, 
MITRE ATT\&CK\footnote{MITRE ATT\&CK is a knowledge base of adversarial tactics and techniques based on real-world observations, https://attack.mitre.org/.}, 
as well as AI-specific ones, 
MITRE ATLAS\footnote{MITRE ATLAS is a knowledge base of adversarial tactics and techniques targeting AI-enabled systems, https://atlas.mitre.org/.}, 
OWASP Top 10 ML\footnote{The OWASP Machine Learning Security Top 10 identifies the most important security risks in machine learning systems, https://owasp.org/www-project-machine-learning-security-top-10/.}, 
and OWASP Top 10 LLM\footnote{The OWASP Top 10 for Large Language Model Applications identifies major security risks in LLM-based applications, https://owasp.org/www-project-top-10-for-large-language-model-applications/.} \cite{jedrzejewski2025thremolia}. The work is conducted in collaboration with Ericsson to evaluate and apply a new threat modeling approach for telecommunication environments, including 5G core networks, network management, and Business and Operation support systems. In particular, the longer-term goal is to validate the effectiveness of LLM-assisted threat modeling in supporting security analysis for telecom-grade, cloud-native infrastructure and software-intensive network systems.
MTMT is a widely used STRIDE-based tool grounded in Microsoft SDL practices \cite{scandariato2015descriptive, microsofttmt}. The experiment uses two intentionally vulnerable applications: AzureGoat \cite{azuregoat} and Vulnerable Bank Application (VulnBank) \cite{vulnbank}. The contributions of the paper are threefold: (i) a vulnerability-grounded validation protocol for threat modeling, (ii) a comparative experiment on two vulnerable applications, and (iii) evidence on the threat coverage achieved by ThreMoLIA and MTMT. 

The remainder of this paper is organized as follows. Section II summarizes related work on validating threat modeling results and positions the proposed vulnerability-grounded approach with respect to expert-based baselines. Section III describes ThreMoLIA and the MTMT, along with the two vulnerable benchmark applications used in the study. Section IV presents the experimental design, including the threat modeling procedure and the coverage-based validation method. Section IV reports the results for AzureGoat and VulnBank and compares their vulnerability coverage. Section V discusses the findings, limitations, and implications for future validation on telecommunications infrastructure. Finally, Section VI concludes the paper.

Validation of the tool's output and corresponding metrics, such as precision, error count, and coverage of threat frameworks, will be discussed in the forthcoming paper within the ThreMoLIA project.

\section{Background and Related Work}
Threat modeling usually starts from an abstraction of the system, typically a Data Flow Diagram (DFD), and then applies a threat framework such as STRIDE to identify candidate threats \cite{shostack2014threat, scandariato2015descriptive}.

ThreMoLIA extends this conventional DFD-based approach with LLM support. The approach combines retrieval-augmented generation (RAG) used to store threat frameworks, data aggregation from design artifacts and prior threat models, prompt construction, and a quality-assurance step to generate and assess LLM-generated threats \cite{jedrzejewski2025thremolia}. ThreMoLIA is motivated by LLM-integrated applications (added support for AI-related frameworks), yet its workflow remains compatible with conventional DFD- and STRIDE-based threat modeling.

Validation of threat modeling approaches has typically relied on human comparison. In practice, such reference models often act as a ``golden standard model'' authored by a subject-matter expert. Wuyts \emph{et al.} evaluated the LINDDUN privacy threat modeling methodology by measuring correctness, completeness, productivity, and ease of use, and compared the results to a panel of privacy experts \cite{wuyts2014linddun}. More recently, Van Landuyt \emph{et al.} constructed expert and novice human baselines to benchmark LLM-based threat modeling tools \cite{vanlanduyt2026benchmark}. 
Recent surveys compare the capabilities and modeling assumptions of automated tools such as MTMT and OWASP Threat Dragon, which aim to reduce human omissions by systematically deriving threats from static threat libraries \cite{granata2024automated, microsofttmt, threatdragon}.  Other empirical work compares threat modeling techniques based on productivity and prioritization behavior rather than on direct linkage to known vulnerabilities \cite{tuma2021matter}. These studies are valuable, but they still depend on human baselines or task interpretations. Our experiment complements them with a different oracle: the known weaknesses and vulnerabilities of intentionally vulnerable applications. While vulnerability-grounded benchmarks exist for code-level security tools (e.g., OWASP Benchmark), applying this approach to design-level threat modeling, where the abstraction gap between threats and vulnerabilities is larger, has not been systematically explored.

\section{Research Design}
The study addresses the following research question: \emph{To what extent can the threats produced by given threat modeling tools supplied with architecture, DFD, and general description as input cover the known vulnerabilities in analyzed applications?}

\subsection{Objects of Analysis}
AzureGoat is a vulnerable-by-design Azure infrastructure that includes web application vulnerabilities and cloud misconfigurations across services such as App Functions, CosmosDB, Storage Accounts, and identities \cite{azuregoat}. VulnBank is a deliberately vulnerable banking application for practicing security testing of web, API, and AI-integrated applications; its attack surface includes classic web flaws as well as AI customer support vulnerabilities \cite{vulnbank}. In the scored experiment, the ground-truth set contained 7 vulnerabilities for AzureGoat and 63 vulnerabilities for VulnBank, as captured in the evaluation workbook.

\subsection{Procedure and Metric}
For each application, an architecture diagram and a DFD were prepared and supplied to both tools. The known-vulnerability list was not used during threat modeling; it was used only during scoring. ThreMoLIA was evaluated on AzureGoat with 0-shot (100\% coverage was achieved with 0-shot) and with $n$-shot on VulnBank, giving additional information from the project documentation at each subsequent shot, and a merged result combining the one-shot and two-shot outputs. MTMT was executed once per application on the same system representation. The GPT-5  model via OpenAI API was used in the experiments which is one of the most advanced frontier models and widely used in the industry and academia. To reduce the effect of run-to-run variability inherent in LLM-based systems, we ran ThreMoLIA three times for each configuration and based the analysis on aggregated results rather than a single run.

A vulnerability was marked as \emph{covered} by an expert if at least one modeled threat semantically described the vulnerability or the condition enabling its exploitation. The main outcome measure is threat coverage:
\begin{equation}
Coverage = \frac{V_{covered}}{V_{known}}
\end{equation}
where $V_{known}$ is the set of known vulnerabilities in the application and $V_{covered}$ is the subset covered by at least one elicited threat. The metric is recall-like; it rewards the discovery of actual weaknesses and vulnerabilities, but it does not penalize false positives. It should be noted that false positives resulting from the LLM's hallucinations are automatically identified during the LLM-powered validation step and removed from the report. The rest of the generated threats that match the selected threat framework, such as STRIDE, but that have no corresponding vulnerabilities in the apps, are still considered valid due to the nature of the threat modeling exercise, which has no theoretical upper limit for the number of modeled valid threats. Therefore, false positives are not considered here. The details of the validation approach will be examined in a forthcoming paper.

\section{Results}
\begin{figure}
    \centering
    \includegraphics[width=0.75\linewidth]{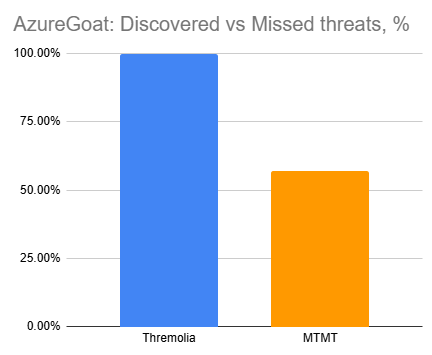}
    \caption{Discovered vulnerabilities in AzureGoat by ThreMoLIA and MTMT}
    \label{fig:AzureGoatCoverage}
\end{figure}

\begin{figure}
    \centering
    \includegraphics[width=1\linewidth]{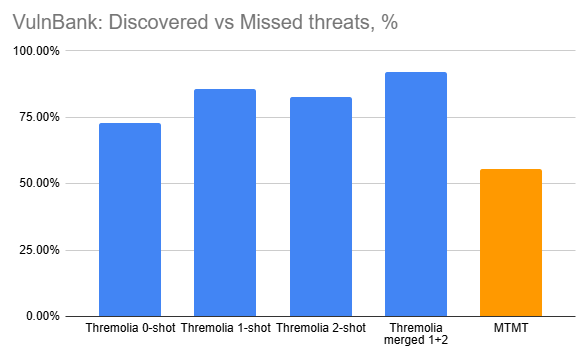}
    \caption{Discovered vulnerabilities in VulnBank by ThreMoLIA and MTMT}
    \label{fig:VulnBankCoverage}
\end{figure}

Table~\ref{tab:coverage} summarizes the results. ThreMoLIA achieved a higher coverage than MTMT on both applications. On AzureGoat~\ref{fig:AzureGoatCoverage}, ThreMoLIA covered all 7 known vulnerabilities, while MTMT covered 4. On VulnBank~\ref{fig:VulnBankCoverage}, even the zero-shot (only a data flow diagram was provided) ThreMoLIA run covered more vulnerabilities than MTMT (46 vs. 35). Coverage improved further with providing additional pieces of documentation, reaching 54/63 for one-shot, 52/63 for two-shot, and 58/63 in the merged case of one-shot and two-shot.

\begin{table}[t]
\caption{Threat coverage against known vulnerabilities.}
\label{tab:coverage}
\centering
\footnotesize
\begin{tabular}{llcc}
\toprule
Application & Tool/configuration & Covered & Coverage \\
\midrule
AzureGoat & ThreMoLIA & 7/7 & 100.0\% \\
AzureGoat & MTMT & 4/7 & 57.1\% \\
\hline
VulnBank & ThreMoLIA 0-shot & 46/63 & 73.0\% \\
VulnBank & ThreMoLIA 1-shot & 54/63 & 85.7\% \\
VulnBank & ThreMoLIA 2-shot & 52/63 & 82.5\% \\
VulnBank & ThreMoLIA merged 1+2 & 58/63 & 92.1\% \\
VulnBank & MTMT & 35/63 & 55.6\% \\
\bottomrule
\end{tabular}
\end{table}

For VulnBank, MTMT performed well on conventional session management and client-side flaws, achieving full coverage in both categories, but it missed all 10 AI customer support vulnerabilities. In contrast, the best ThreMoLIA configurations covered all 10 AI vulnerabilities and all 9 authentication/authorization vulnerabilities. This suggests that MTMT remains useful for classical STRIDE-style issues, whereas ThreMoLIA is better suited to mixed application landscapes that combine conventional and AI-related attack surfaces.

If the best-performing ThreMoLIA configuration is taken as representative, overall coverage across both applications is 65/70 vulnerabilities (92.9\%), compared with 39/70 (55.7\%) for MTMT. At the same time, the difference between zero-shot, one-shot, and two-shot runs shows that prompt strategy matters. The merged result improved over either single-prompt run, indicating that multiple LLM-assisted threat modeling runs can reveal more threats.

\section{Discussion}
The results support the usefulness of vulnerable applications as validation benchmarks for threat modeling. Compared with expert-only baselines, this approach offers a measurable alternative, i.e., an oracle: the weaknesses and vulnerabilities being scored are known to exist in the target system. This does not eliminate all subjectivity, because the mapping between a generated threat and a vulnerability still requires analyst judgment, as a threat and a vulnerability may have different text descriptions. However, it shifts the evaluation away from \emph{``Did the tool match one specific expert report?''} to \emph{``Did the tool cover weaknesses and vulnerabilities that are actually present?''}

The experiment also shows a clear difference between the two tools. MTMT is rooted in a mature DFD-and-STRIDE workflow and remains effective for standard architectural concerns \cite{scandariato2015descriptive}. However, ThreMoLIA achieved substantially higher coverage, especially on VulnBank, whose vulnerability set mixes authentication, business-logic, and AI-related weaknesses. This result is consistent with the design goals of ThreMoLIA, which explicitly combine diagram understanding with broader contextual knowledge and tool-supported quality assurance \cite{jedrzejewski2025thremolia}. The effect of prompting on VulnBank also aligns with recent benchmark findings that no single prompting strategy is uniformly best for threat modeling \cite{vanlanduyt2026benchmark}.

\paragraph{Limitations}
First, the coverage is only one dimension of quality; a tool may achieve high recall while still generating irrelevant threats. Second, the merged ThreMoLIA result combines two prompted runs and is therefore not strictly cost-equivalent to a single deterministic MTMT run. Third, only two applications were used. Fourth, GPT-5 model could be trained on vulnerable applications' data. To mitigate this risk, we renamed both applications when feeding the DFDs and descriptions to the LLM. Future work should extend this design to include more vulnerable applications and additional metrics, such as precision, analyst effort, and time.

\section{Conclusion}
This paper presented a brief comparative study of validating threat modeling results using vulnerable test applications. Instead of relying solely on human experts' assessments, we used the known-vulnerability sets of AzureGoat and VulnBank as scoring oracles to measure threat coverage. Across both applications, ThreMoLIA consistently identified more actual vulnerabilities than MTMT, with the largest difference observed on VulnBank's mixed web/API/AI attack surface. The results indicate that vulnerability-grounded validation is a practical complement to expert baselines and can help assess how well threat modeling tools discover weaknesses and vulnerabilities that truly exist in the system under analysis. Moreover, these deliberately vulnerable open-source applications could end up in the testing benchmark used to evaluate LLM-assisted threat modeling tools. \textbf{As a next step, the same methodology will be further validated on telecommunications environments, including cloud-native network functions and 5G core infrastructure, where architecture-driven threat modeling is especially relevant. ThreMoLIA tool is also planned to be released as open source.}

\section*{Acknowledgment}
We would like to acknowledge that this work was supported by the KKS foundation through the SERT Research Profile project (research profile grant 2018/010) at Blekinge Institute of Technology and the Threat Modeling for LLM-Integrated applications(ThreMoLIA) Research Project supported by Vinnova (Sweden’s Innovation Agency) (Diarienummer 2024-00659).

\bibliographystyle{IEEEtran}
\bibliography{references}


\appendices
\section{Appendix: Detailed Vulnerability Coverage Tables}

\noindent\begin{minipage}[t]{\columnwidth}
\captionsetup{type=table}
\caption{AzureGoat detailed threat coverage.}
\label{tab:app-azuregoat}
\centering
\scriptsize
\setlength{\tabcolsep}{3.5pt}
\renewcommand{\arraystretch}{0.93}
\begin{tabular}{p{0.66\columnwidth}cc}
\toprule
Vulnerability & ThreMoLIA & MTMT \\
\midrule
XSS & \textbf{Discovered} & Missed \\
NoSQL Injection & \textbf{Discovered} & \textbf{Discovered} \\
Insecure Direct Object Reference & \textbf{Discovered} & Missed \\
Server-Side Request Forgery on App Function Environment & \textbf{Discovered} & Missed \\
Sensitive Data Exposure and Password Reset & \textbf{Discovered} & \textbf{Discovered} \\
Storage Account Misconfigurations & \textbf{Discovered} & \textbf{Discovered} \\
Identity Misconfigurations & \textbf{Discovered} & \textbf{Discovered} \\
\bottomrule
\end{tabular}

\vspace{0.8em}

\captionsetup{type=table}
\caption{VulnBank detailed threat coverage.}
\footnotesize T0 = ThreMoLIA 0-shot, T1 = 1-shot, T2 = 2-shot, TM = merged, D = discovered, M = missed.
\label{tab:app-vulnbank}
\centering
{\tiny
\setlength{\tabcolsep}{2.1pt}
\renewcommand{\arraystretch}{0.84}
\begin{tabular}{p{0.66\columnwidth}ccccc}
\toprule
\textbf{Vulnerability} & \textbf{T0} & \textbf{T1} & \textbf{T2} & \textbf{TM} & \textbf{MTMT} \\
\midrule
SQL Injection in login & D & D & D & D & D \\
Weak JWT implementation & D & D & D & D & D \\
Broken object level authorization (BOLA) & D & D & D & D & M \\
Broken object property level authorization (BOPLA) & D & D & D & D & M \\
Mass Assignment \& Excessive Data Exposure & D & D & D & D & M \\
Weak password reset mechanism (3-digit PIN) & M & M & D & D & D \\
Token stored in localStorage & D & D & D & D & D \\
No server-side token invalidation & M & D & D & D & D \\
No session expiration & M & D & D & D & D \\
Information disclosure & D & D & D & D & D \\
Sensitive data exposure & D & D & D & D & D \\
Plaintext password storage & M & M & M & M & M \\
SQL injection points & D & D & D & D & D \\
Debug information exposure & D & D & D & D & M \\
Detailed error messages exposed & D & D & D & D & D \\
No amount validation & D & D & D & D & D \\
Negative amount transfers possible & D & D & D & D & M \\
No transaction limits & M & M & M & M & D \\
Race conditions in transfers and balance updates & M & M & D & D & D \\
Transaction history information disclosure & D & D & D & D & M \\
No validation on recipient accounts & D & D & M & D & D \\
Unrestricted file upload & D & D & D & D & M \\
Path traversal vulnerabilities & D & D & D & D & M \\
No file type validation & D & D & D & D & D \\
Directory traversal & D & D & D & D & D \\
No file size limits & D & D & D & D & D \\
Unsafe file naming & D & D & D & D & M \\
Token vulnerabilities & D & D & D & D & D \\
No session expiration & M & D & D & D & D \\
Weak secret keys & D & D & D & D & D \\
Token exposure in URLs & D & D & D & D & D \\
Cross Site Scripting (XSS) & D & D & D & D & D \\
Cross Site Request Forgery (CSRF) & D & D & D & D & D \\
Insecure direct object references & D & D & D & D & D \\
No rate limiting & D & D & D & D & D \\
Mass Assignment in card limit updates & D & D & D & D & M \\
Predictable card number generation & M & M & M & M & M \\
Plaintext storage of card details & M & D & M & D & M \\
No validation on card limits & M & D & D & D & D \\
BOLA in card operations & D & D & D & D & M \\
Race conditions in balance updates & M & M & D & D & D \\
Card detail information disclosure & D & D & D & D & D \\
No transaction verification & D & D & M & D & D \\
Lack of card activity monitoring & M & D & M & D & D \\
No validation on payment amounts & D & D & D & D & D \\
SQL injection in biller queries & M & D & D & D & D \\
Information disclosure in payment history & D & D & D & D & D \\
Predictable reference numbers & M & M & M & M & M \\
Transaction history exposure & D & D & D & D & M \\
No validation on biller accounts & D & D & M & D & D \\
Race conditions in payment processing & M & M & D & D & D \\
BOLA in payment history access & D & D & D & D & M \\
Missing payment limits & M & M & M & M & M \\
Prompt Injection (CWE-77) & D & D & D & D & M \\
AI-based Information Disclosure (CWE-200) & D & D & D & D & M \\
Broken Authorization in AI context (CWE-862) & D & D & D & D & M \\
AI System Information Exposure (CWE-209) & D & D & D & D & M \\
Insufficient Input Validation for AI prompts (CWE-20) & D & D & D & D & M \\
Direct Database Access through AI manipulation & M & D & M & D & M \\
AI Role Override attacks & D & D & D & D & M \\
Context Injection vulnerabilities & D & D & D & D & M \\
AI-assisted unauthorized data access & D & D & D & D & M \\
Exposed AI system prompts and configurations & D & D & D & D & M \\
\bottomrule
\end{tabular}
}
\end{minipage}

\end{document}